\documentclass[aps,prl,showpacs,twocolumn,floats,psfig,superscriptaddress]{revtex4}
\usepackage{amssymb}
\usepackage{amsbsy}
\usepackage{amsmath}
\usepackage{epsf}
\usepackage{graphicx}

\def\nn{\nonumber}

\def\be{\begin{equation}}
\def\ee{\end{equation}}
\def\bea{\begin{eqnarray}}
\def\eea{\end{eqnarray}}
\def\ba{\begin{array}}
\def\ea{\end{array}}

\begin{document}

\title{Graphene - a nearly perfect fluid}
\author{Markus M\"uller}
\affiliation{The Abdus Salam International Center for Theoretical Physics, Strada Costiera 11, 34014 Trieste, Italy
}
\author{J\"{o}rg Schmalian}
\affiliation{Ames Laboratory and Department of Physics and Astronomy, Iowa State
University, Ames, IA 50011, USA}
\author{Lars Fritz}
\affiliation{Department of Physics, Harvard University, Cambridge MA 02138, USA }
\date{\today\\
}
\pacs{67.90.+z,71.10.-w,73.23.-b,81.05.Uw}

\begin{abstract}
Hydrodynamics and collision dominated transport are crucial to understand
the slow dynamics of many correlated quantum liquids. The ratio $\eta/s$ of
the shear viscosity $\eta$ to the entropy density $s$ is uniquely suited to
determine how strongly the excitations in a quantum fluid interact. We
determine $\eta/s$ in clean undoped graphene using a quantum kinetic theory.
As a result of the quantum criticality of this system the ratio is smaller
than in many other correlated quantum liquids and, interestingly, comes
close to a lower bound conjectured in the context of the quark gluon plasma.
We discuss possible consequences of the low viscosity, including
pre-turbulent current flow.
\end{abstract}

\maketitle

Graphene~\cite{Novoselov04,Novoselov05}, attracts a lot of attention due to
the massless relativistic dispersion of its quasiparticles and their high
mobility.
Recently, it was shown that this material offers a unique opportunity to
observe transport properties of a plasma of ultrarelativistic particles at
moderately high temperatures~\cite{Fritz08}. Undoped graphene is located at
a special point in parameter space where the Fermi surface shrinks to two
points, and in many respects it behaves similarly as other systems close to
more complex quantum critical points~\cite{Sheehy07}. Due to its massless
Dirac particles
graphene also shares interesting properties with the ultrarelativistic quark
gluon plasma. The latter, surprisingly, has an unexpectedly low shear
viscosity, as was observed in the dense matter balls created at the
relativistic heavy ion collider RHIC~\cite{Shuryak03}. We show here that an
analogous property can be found in undoped graphene, reflecting its quantum
criticality.

\begin{figure}[b]
\centerline{\includegraphics[width=0.42\textwidth]{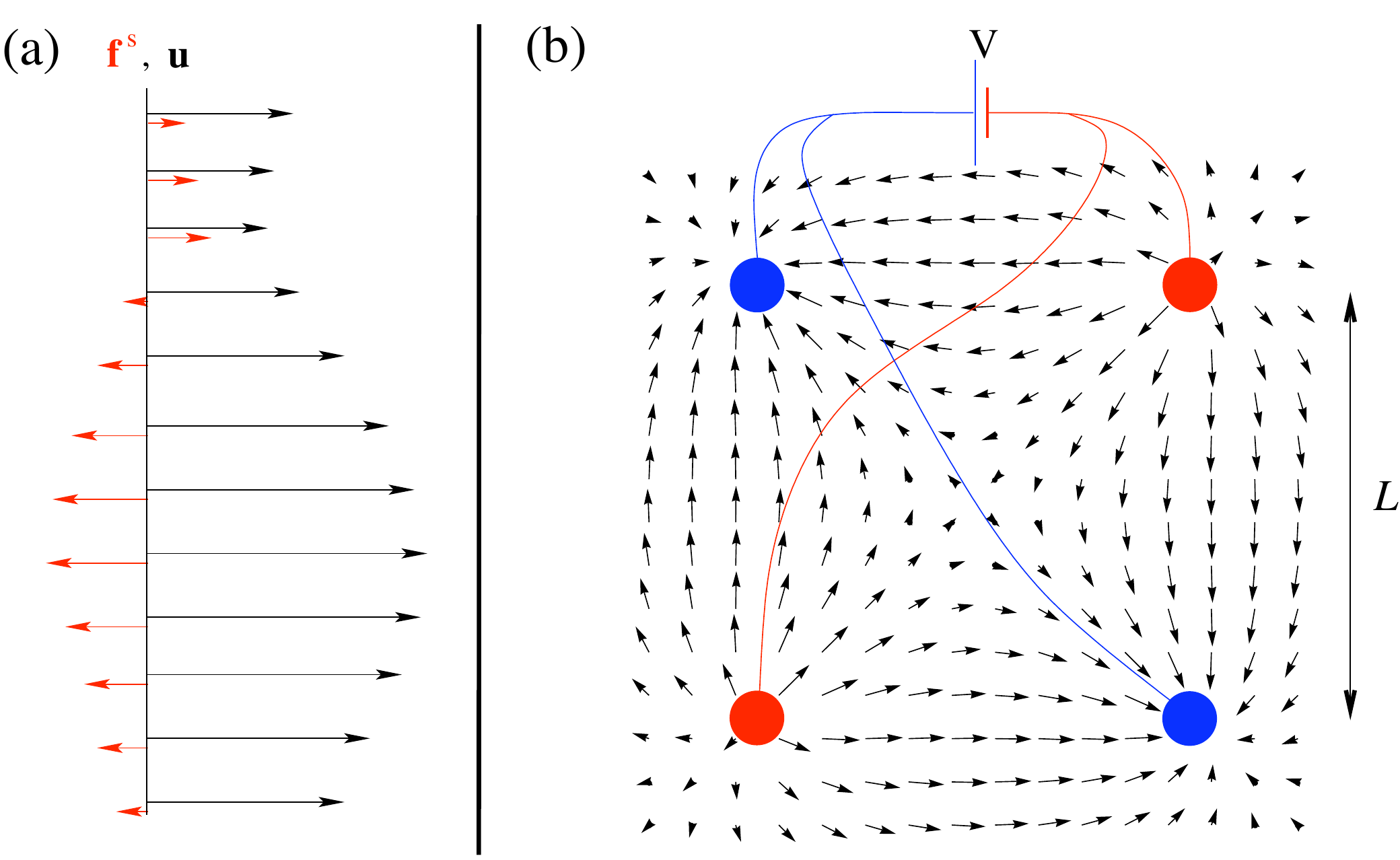}} 
\caption{(a) Velocity profile $\mathbf{u}$ and associated Stokes force
density $\mathbf{f^s}=\protect\eta \protect\nabla^2 \mathbf{u}$
counteracting the current flow. {(b)} Inhomogeneous current flow expected in
a four-contact geometry with split source and drain contacts held at voltage
$\pm V/2$. In the absence of viscous and other nonlocal effects, the current
would be proportional to the applied voltage $V$, independent of the
distance $L$ between the contacts. Viscous effects diminish the current as $L
$ decreases. }
\label{Fig:flowfield}
\end{figure}

The shear viscosity $\eta$ measures the resistance of a fluid to
establishing transverse velocity gradients, see Fig.~\ref{Fig:flowfield}.
The smaller the viscosity,
the higher the tendency to turbulent flow dynamics. Viscosity, similarly as
resistivity in a conductor, leads to entropy production by degrading
inhomogeneities in the velocity field. While \textit{ideal fluids} with $%
\eta =0$ cannot exist, it is interesting to seek for \textit{perfect fluids}
which come as close as possible to this ideal.

Viscosity has the units of $\hbar n$ where $n$ is some density. To quantify
the magnitude of the viscosity, it is natural to compare $\eta/\hbar $ to
the density of thermal excitations, $n_{\mathrm{th}}$, which can be
estimated by the entropy density, $s\sim k_{B}n_{\mathrm{th}}$. Motivated by
the nearly perfect fluid behavior seen in the RHIC experiments, Kovtun et
al. have recently postulated a lower bound for the ratio of $\eta $ and $s$
for a wide class of systems~\cite{Kovtun03}:
\begin{equation}
\eta /s\geq \frac{1}{4\pi }\frac{\hbar }{k_{B}}.  \label{bound}
\end{equation}
Equality was obtained for an infinitely strongly coupled conformal field
theory by mapping it to weakly coupled gravity using the AdS-CFT
correspondence. While examples violating the bound (\ref{bound})were found
(see~\cite{counterexample}), the existence of some lower bound with $%
k_{B}\eta /\hbar s$ of order unity for a given family of fluids is not
unexpected. It is analogous to the Mott-Ioffe-Regel limit for the minimum
conductivity of poor metals~\cite{Ioffe60,Mott72}, and to the saturation of
the relaxation rate at $\tau_{\mathrm{rel}}^{-1}=k_{B}T/\hbar \cdot \mathcal{%
O}(1)$ close to strongly coupled quantum critical points~\cite{ssbook}. In
all these cases an exhaustion of scattering channels and a saturation of the
kinetic coefficients occurs, once the mean free path becomes comparable to
the interparticle distance. It follows, that the ratio $\eta /s$ is a unique
indicator for how strongly the excitations in a fluid interact.

While effects due to electron-electron interactions only amount to very
small additive corrections to the conductivity $\sigma \left( \omega
,T\right) $ in the collisionless optical regime $\hbar \omega \gg k_{B}T$~%
\cite{Sheehy07,Herbut08,Fritz08}, collisions are crucial in the opposite
regime $\hbar \omega \ll k_{B}T$~\cite{Fritz08,Kashuba08}. There they
establish local equilibrium, the remaining low frequency dynamics being
governed by hydrodynamics, i.e., by the slow diffusion of the densities of
globally conserved quantities. Clearly, a hydrodynamic description works
best when collisions are frequent and the fluid is strongly correlated.
Transport in such a regime reveals information about the nature of the
excitations. Examples ranging from the two He-isotopes~\cite%
{Abrikosov59,Khalat}, cold atomic gases~\cite{Bruun07,Schafer07} and
electrons in semiconductors and metals~\cite{Steinberg58,Vignale97},
extending all the way to the high energy regime of the quark gluon plasma~%
\cite{Shuryak03} and matter in the early universe~\cite{Weinberg71}.

In graphene, for energies below a few electron volts, the electronic
properties are governed by the Hamiltonian
\begin{equation}
H=\sum\limits_{l}v\widehat{\mathbf{p}}_{l}\mathbf{\cdot \sigma +}\frac{1}{2}%
\sum\limits_{l, l^{\prime }}\frac{e^{2}}{\epsilon \left\vert \mathbf{r}_{l}-%
\mathbf{r}_{l^{\prime }}\right\vert },  \label{Ham}
\end{equation}
with the Fermi velocity $v\simeq 10^{8}\mathrm{cm/}$\textrm{s}~\cite%
{Novoselov05}. $\widehat{\mathbf{p}}_{l}=-i\hbar \mathbf{\nabla}_{\mathbf{r}%
_{l}}$ is the momentum operator, $l=1,...,N$ labels the $N=4$ spin and
valley (2 Fermi points) degrees of freedom, and $\mathbf{\sigma }=\left(
\sigma _{x},\sigma _{y}\right) $ are the Pauli matrices acting in the space
of the two sub-lattices of the honeycomb lattice structure.
Without the Coulomb interaction, Eq.~\eqref{Ham} is the Hamiltonian of $N$
species of free massless Dirac particles~\cite{Wallace47}. The strength of
the Coulomb interaction is
characterized by the effective fine structure constant
$\alpha =\frac{e^{2}}{ \epsilon \hbar v} 
\simeq 2.2/\epsilon$, which is not small 
for realistic values of the substrate dielectric constant $\epsilon$.
Key for an understanding of clean, undoped graphene is the fact that it is a
`quantum critical' system with marginally irrelevant Coulomb interactions
which renormalize logarithmically to zero~\cite%
{Gonzalez99,Ye98,gorbar,Son07,Sheehy07,Herbut08,Fritz08}. This quantum
critical behavior in undoped graphene is responsible for the distinctly
different behavior of the collision-free and collision-dominated frequency
regimes~\cite{Damle97}.

Collision-dominated transport can most efficiently be addressed by solving
the Boltzmann transport equation
\begin{equation}
\left( \frac{\partial }{\partial t}+\frac{1}{\hbar }\frac{\partial
\varepsilon _{\mathbf{k}\lambda }}{\partial \mathbf{k}}\cdot \mathbf{\nabla }%
_{\mathbf{x}}-\frac{1}{\hbar }\frac{\partial \varepsilon _{\mathbf{k}\lambda
}}{\partial \mathbf{x}}\cdot \mathbf{\nabla }_{\mathbf{k}}\right) f=-%
\mathcal{J}_{\mathrm{coll}}\left[ f\right]   \label{Boltzmann}
\end{equation}%
for the quasiparticle distribution function $f=f_{\lambda }\left( \mathbf{k},%
\mathbf{x},t\right) $, which depends on momentum $\mathbf{k}$, position $%
\mathbf{x}$, time $t$ and band index $\lambda =\pm $ (labelling upper and
lower parts of the Dirac cones centered at the two Fermi points).
Eq.~\eqref{Boltzmann} can be derived from a nonequilibrium quantum many body
approach, yielding the collision integral $\mathcal{J}_{\mathrm{coll}}$
in terms of the Coulomb interaction~\cite{Fritz08}. The moments of the
distribution function yield the conservation laws for the charge density,
$\partial _{t}\rho +\mathbf{\nabla }\cdot \mathbf{j}=0,$
 the momentum density,
\begin{equation}
\frac{w}{v^{2}}\left[ \partial _{t}\mathbf{u}+\left( \mathbf{u}\cdot \mathbf{%
\nabla }\right) \mathbf{u}\right] +\mathbf{\nabla }p+\frac{\partial _{t}p}{%
v^{2}}\mathbf{u}+\mathbf{f}^{s}=0,  \label{mom}
\end{equation}%
and the energy density $\varepsilon $. $\mathbf{j}$\ is the current density
and $\mathbf{u}$ the velocity field of the fluid with enthalpy density $%
w=\varepsilon +p$ where $p$ is the pressure. For undoped graphene the
Gibbs-Duhem relation implies furthermore $w=Ts$. Eq.~(\ref{mom}) is the
Navier-Stokes equation for graphene, derived under the assumption $%
\left\vert \mathbf{u}\right\vert \ll v_{F}$. Compared to non-relativistic
hydrodynamics there is an extra relativistic term $\propto \partial _{t}p$,
but at low frequencies its effect is small. The Stokes force ${f}%
_{j}^{s}=\partial _{i}{T}_{ij}=\eta \nabla ^{2}{u}_{j}$ is determined by the
leading dissipative contribution to the stress tensor:
\begin{equation}
{T}_{ij}=\eta X_{ij}+\zeta \delta _{ij}\mathbf{\nabla }\cdot {\mathbf{u}},
\label{lin}
\end{equation}%
in an expansion in gradients of ${\mathbf{u}}$. Here $X_{ij}=\partial
u_{i}/\partial x_{j}+\partial u_{j}/\partial x_{i}-\delta _{ij}\mathbf{%
\nabla }\cdot \mathbf{u}$ corresponds to a pure shear flow while the second
term is a volume compression. $\eta $ and $\zeta $ are the shear and bulk
viscosity, respectively.

In what follows we include all contributions to $\mathcal{J}_{\mathrm{coll}}$
up to order $\alpha ^{2}$, keeping in mind that $\alpha $ flows to zero as $%
T\rightarrow 0$. Assuming a slowly varying, divergence free velocity field ${%
\mathbf{u}}(\mathbf{r})$, we determine the shear viscosity by computing the
stress tensor in linear response. Close to equilibrium, an inhomogeneous
flow field constitutes a driving term in Eq.~\eqref{Boltzmann} of the form $%
\mathbf{v}_{\mathbf{k},\lambda }\cdot \mathbf{\ \nabla }_{\mathbf{x}%
}f_{\lambda }=\sum_{ij}\phi _{ij}X_{ji}$ with
\begin{equation}
\phi _{ij}\left( \mathbf{k},\lambda \right) =\frac{\varepsilon _{\mathbf{k}%
\lambda }}{k_{B}T}\frac{e^{\beta \varepsilon _{\mathbf{k}\lambda }}}{%
2^{3/2}\left( e^{\beta \varepsilon _{\mathbf{k}\lambda }}+1\right) ^{2}}%
I_{ij}\left( \mathbf{k}\right) .  \label{phi}
\end{equation}%
Here, $\mathbf{v}_{\mathbf{k},\lambda }=\mathbf{\nabla }_{\mathbf{k}%
}\varepsilon _{\mathbf{k}\lambda }/\hbar $ is the velocity of quasiparticles
with energy $\varepsilon _{\mathbf{k}\lambda }$, $\beta =1/k_{B}T$ and $%
I_{ij}\left( \mathbf{k}\right) =\sqrt{2}\left( \frac{k_{i}k_{j}}{k^{2}}-%
\frac{1}{2}\delta _{ij}\right) $. In linear response, the distribution
function 
can be parametrized as:
\begin{eqnarray}
f_{\lambda }\left( \mathbf{k},t\right)  &=&\frac{1}{\exp \left( \beta \left[\varepsilon _{\mathbf{k}\lambda }-\hbar \sum_{ij}X_{ij}g_{ji}\left( \mathbf{k,}\lambda ,t\right) \ \right] \right) +1}\nn \\
&\approx &f_{\mathrm{eq}}+f_{\mathrm{eq}}(1-f_{\mathrm{eq}})\beta \hbar\sum_{ij}X_{ij}g_{ji},  \label{offeq_f}
\end{eqnarray}%
where $f_{\mathrm{eq}}=f_{\lambda }(\mathbf{k})|_{g_{ij}=0}$. Linearizing
the Boltzmann equation in the zero frequency limit, it can be cast into an
operator formulation: $\left\vert \phi \right\rangle =\mathcal{C}\left\vert g\right\rangle $
~\cite{yaffe00,ziman}.
The operator $\mathcal{C}$ is Hermitian with respect to the inner product $%
\left\langle a|b\right\rangle =\left( 8\pi ^{2}\right) ^{-1}\sum_{ij,\lambda
}\int d^{2}k\,a_{ij}\left( \mathbf{k,}\lambda \right) b_{ji}\left( \mathbf{k}%
,\lambda \right) $. The $g_{ij}\left( \mathbf{k,}\lambda \right) $
parametrize the non-equilibrium distribution function and are obtained by
inverting the operator $\mathcal{C}$. Using them to express the stress
tensor and comparing with Eq.~(\ref{lin}) one obtains the shear viscosity:
\begin{equation}
\eta =\frac{N\left( k_{B}T\right) ^{2}}{\sqrt{2}\hbar v^{2}}\left\langle
\phi \left\vert \mathcal{C}^{-1}\right\vert \phi \right\rangle \,.
\label{shear}
\end{equation}%
The dominant contribution to $\eta $ comes from the smallest eigenvalues of $%
\mathcal{C}$ restricted to functions given by Eq.~\eqref{phi}. The inversion
of the collision operator can be a formidable problem and usually requires a
numerical solution. The problem simplifies, however, in two dimensions where
the amplitude for collinear scattering processes (involving quasiparticles
with identical velocity vector) is logarithmically divergent. Screening
effects of higher order in $\alpha $ and lifetime effects cut off this
divergence in the infrared at transverse momenta of order $\alpha T/v$~\cite%
{Kashuba08,Fritz08}. To logarithmic accuracy in $\alpha $, we can thus
consider collinear scattering processes only. The corresponding restricted
operator possesses three zero modes:
\begin{eqnarray}
g_{ij}^{\left( n\right) }\left( \mathbf{k,}\lambda \right)
&=&c^{(n)}\,I_{ij}\left( \mathbf{k}\right) ,  \notag \\
g_{ij}^{\left( \mathrm{\chi }\right) }\left( \mathbf{k,}\lambda \right)
&=&c^{(\chi )}\,\lambda I_{ij}\left( \mathbf{k}\right) ,  \notag \\
g_{ij}^{\left( E\right) }\left( \mathbf{k},\lambda \right)
&=&c^{(E)}\,\lambda |{\mathbf{k}}|I_{ij}\left( \mathbf{k}\right) ,
\label{modes}
\end{eqnarray}%
which reflect the conservation of charge $n$, chirality $\chi $ (the total
number of particles and holes) and energy $E$ in collinear processes. This
conservation is exact for the modes $g^{(n,E)}$, while it holds only to
lowest order in $\alpha $ for $g^{(\chi )}$, being due to kinetic
constraints on the two-body scattering of massless Dirac particles (see~\cite%
{Foster2009} for a related discussion). Eq.~(\ref{offeq_f}) shows that these
modes correspond to distribution functions which, when restricted to
quasiparticles with identical velocity $\mathbf{v}=v\,\mathbf{e}$, reduce to
equilibria with direction dependent parameters $\mu (\mathbf{e}),\varphi (%
\mathbf{e}),T(\mathbf{e})$ conjugate to the conserved quantities.

If there were only collinear scattering processes, the shear viscosity would
be infinite. However, the inclusion of other processes causes $\eta $ to be
finite. Nevertheless, the dominance of collinear scattering allows us - in
leading logarithmic approximation - to invert the operator $\mathcal{C}$
within the Hilbert space spanned by the modes~\eqref{modes}. At zero doping,
a divergence free velocity field does not excite the mode $g_{ij}^{(n)}$,
and the relevant subspace is only two-dimensional. This inversion is easily
done and yields $\left\langle \phi \left\vert \mathcal{C}^{-1}\right\vert
\phi \right\rangle =C_{\eta }$ $2^{-3/2}\alpha ^{-2}$. The remaining
numerical coefficient $C_{\eta }$ stems from the evaluation of the matrix
elements of the full scattering operator in the 2d subspace of $%
g_{ij}^{E,\chi }$ and is expected to be of order unity. We obtain  $%
C_{\eta }\simeq 0.449$, consistent with this expectation. The $%
\alpha ^{-2}$ dependence follows from the fact that the collision
operator ${\cal C}$ is of second order in $\alpha $. The shear
viscosity of graphene finally results as
\begin{equation}
\eta =C_{\eta }\frac{N\left( k_{B}T\right) ^{2}}{4\hbar v^{2}\alpha ^{2}}%
\left[ 1+\mathcal{O}\left( \frac{1}{\log \alpha }\right) \right] ,
\label{shear2}
\end{equation}%
which is the central result of this Letter. It can be
rationalized by using the Fermi liquid result~\cite{Abrikosov59} $\eta
_{\rm FL}\simeq n mv^{2}\tau _{\rm rel}$ with $n\to n_{\rm thermal}\simeq \left( k_{B}T/\hbar v\right) ^{2}$, 
relaxation rate $
\tau _{\rm rel}^{-1}\simeq k_{B}T/\left(\hbar \alpha ^{2}\right) $~\cite{Fritz08} and typical energy $mv^{2}\rightarrow k_{B}T$.  Extending the
analysis beyond the leading approximation by regularizing the logarithmic
divergence in the forward scattering and including more modes $g_{ij}$, we
obtain corrections of relative size $1/\log (1/\alpha )$. For $\alpha =0.1$
they increase the leading result (\ref{shear2}) by only $20\%$.

We need to keep in mind that the quasiparticle are not free, their
dispersion reflecting the renormalization of the velocity
$v\rightarrow v\left[ 1+\frac{\alpha }{4}\log (\Lambda /k)\right] $, where $%
\Lambda $ is an appropriate UV scale. We implement this by a renormalization
group approach combined with scaling laws for physical observables. The
coupling constant evolves as $\alpha \left( T\right) \simeq 4/\log \frac{%
T_{\Lambda }}{T}$ (with $T_{\Lambda }=\frac{\hbar v\Lambda }{k_{B}}$) while
the velocity grows logarithmically $v\left( T\right) =v\alpha /\alpha (T)$.
However, the combination $\left[ \alpha v\right] \left( T\right) $ entering $%
\eta $ does not change under renormalization~\cite{Fisher88}. Thus, Eq.%
\eqref{shear2}  is the correct low temperature result for the renormalized
quasiparticles.

On the other hand, the entropy density of noninteracting graphene, including
renormalization effects, is~\cite{Sheehy07}
\begin{equation}
\ s=\frac{9\zeta \left( 3\right) }{\pi}\,k_B\frac{k_{B}^{2}T^{2}}{ \left(
\hbar v \alpha\right) ^{2}}\alpha^{2}(T).
\end{equation}
The above finally results in the sought ratio:
\begin{equation}
\eta /s=\frac{\hbar }{k_{B}}\frac{C_{\eta }\pi \ }{9\zeta \left( 3\right) }%
\frac{1}{\alpha ^{2}\left( T\right)} \simeq \,0.00815\times \left( \log\frac{%
T_\Lambda}{T}\right) ^{2}.  \label{etaovers}
\end{equation}

\begin{figure}[b]
\centerline{\includegraphics[width=0.42\textwidth]{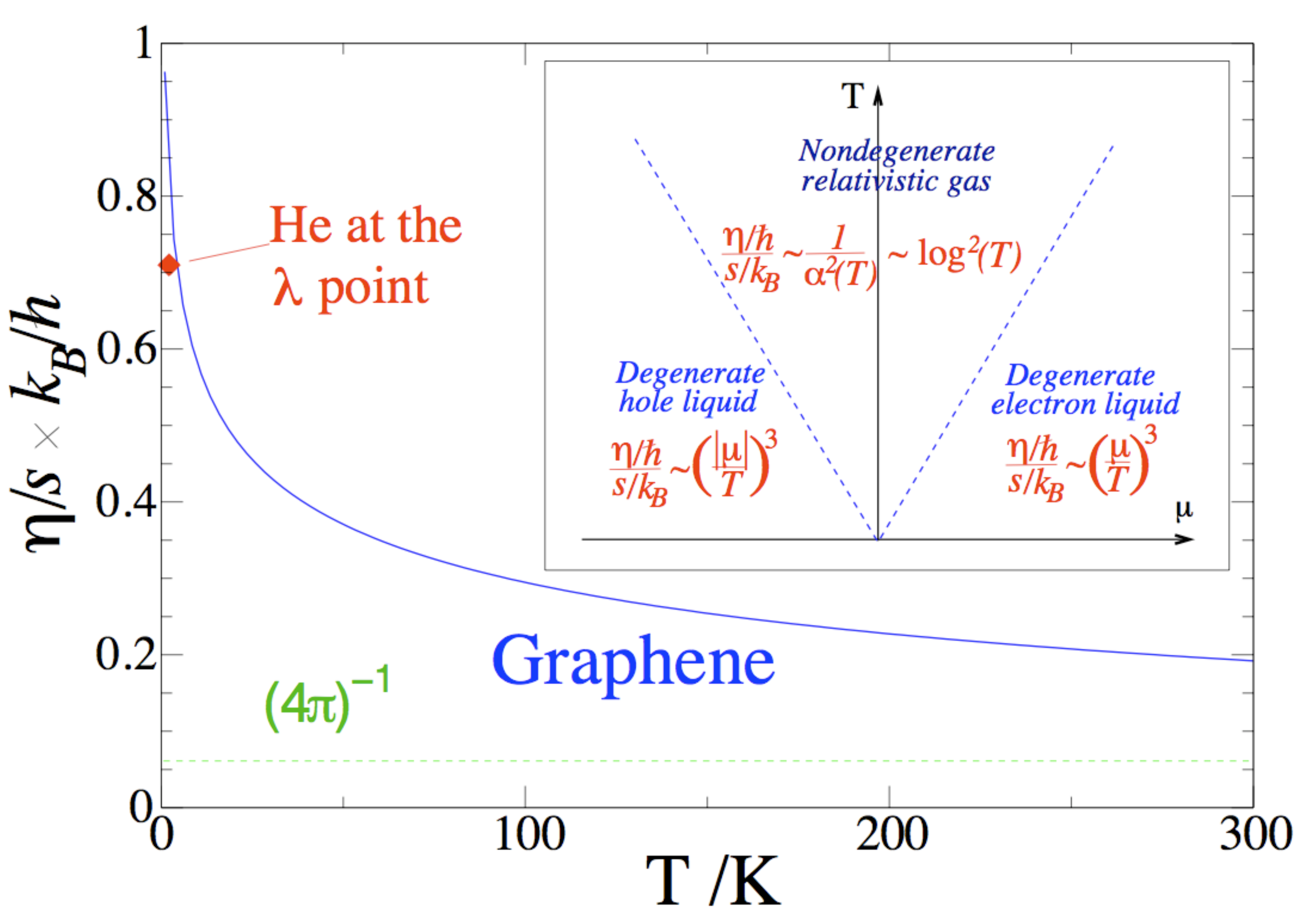}}
\caption{ {Ratio $\protect\eta /s$ in
graphene as a function of $T$.} The UV cut-off was taken to be $T_{\Lambda
}=8.34\cdot 10^{4}\mathrm{K}$ following~\protect\cite{Sheehy07}. In the
undoped, quantum critical system $\protect\eta /s$ is very small in a large
temperature window where the coupling $\protect\alpha (T)$ remains of order $%
O(1)$. The value $1/4\protect\pi $ obtained for some strongly coupled
critical theories is shown for comparison. As shown in the inset, away from
zero doping and quantum criticality ($T<|\protect\mu |$), the viscosity
assumes the behavior of a degenerate Fermi liquid, $\protect\eta /s\sim (|%
\protect\mu |/T)^{3}$.}
\label{Fig:viscovers}
\end{figure}

As $T\rightarrow 0$ the ratio $\eta /s$ grows, a behavior expected for a
weakly interacting system.
However, since $\alpha $ is only marginally irrelevant $\eta /s$ grows only
logarithmically. In contrast, in the regime $T\ll \mu $ of doped graphene
with a finite carrier density $n$, one obtains the usual behavior of a
degenerate Fermi liquid ~\cite{Steinberg58,Abrikosov59} with $\eta \sim
\hbar n(\mu /T)^{2}$and $s\sim k_{B}n\,T/\mu $, in which case $\eta /s\sim
(\hbar /k_{B})\left( \mu /T\right) ^{3}$ diverges much more strongly at low $%
T$, see Fig.~\ref{Fig:viscovers}. Higher order corrections
of the long range Coulomb interaction~\cite{Son07} mainly reduce the
regime where $\alpha \left( T\right) $ decreases logarithmically
(they effectively reduce $T_{\Lambda }$). This further decreases the ratio $\eta /s$. Similarly,
additional short range interactions $g$ yield leading additive corrections
$\propto \alpha g\left( T/T_{\Lambda }\right)$ to the collision
operator. Their effect is small provided that they do not lead to an excitonic insulator and the low energy physics of massless Dirac particles is preserved~\cite{Herbut06}.

Note the small numerical prefactor in (\ref{etaovers}). As shown in Fig.~\ref%
{Fig:viscovers}, it keeps the ratio $\eta /s$ small in a large temperature
regime, where it approaches the value of Eq.~\eqref{bound}. As was shown
recently, cold atoms with diverging scattering length are materials which
also come close to the value \eqref{bound}~\cite{Bruun07,Schafer07}. Our
result shows that, interestingly, graphene has an even smaller ratio $\eta /s
$, thus being an even "more perfect" liquid than those critical systems.
We envision interesting experimental manifestations of viscous effects in
graphene,
especially in the conductance properties of very clean samples. Viscous drag
should result in a decrease of the conductance with the linear size of
geometries such as in Fig.~\ref{Fig:flowfield}, where two spatially
separated contacts take the role of the source and the drain, respectively.
Applying a source-drain bias $V$ to these split contacts induces an
inhomogeneous current flow and corresponding Stokes forces that oppose it.
The smaller the spatial scale $L$, the larger the viscous forces, and hence
one expects an increasing resistance. The latter would be scale-invariant in
the absence of viscous forces and other non-local effects on conductivity~%
\cite{Abanin08}.

The unusually low viscosity in graphene suggests the interesting possibility
of electronic turbulence in this material.
For simplicity we consider fluid velocities small compared to the Fermi
velocity, and analyze the low frequency limit of the Navier-Stokes Eq.~%
\eqref{mom}. Turbulence arises from the nonlinearities $\propto \left(
\mathbf{u}\cdot \mathbf{\nabla }\right) \mathbf{u}$ while dissipation due to
the Stokes forces suppress turbulent flow for large $\eta $. The
dimensionless number determining the relative strength of these two effects
is the Reynolds number:
\begin{equation}
\mathrm{Re}=\frac{s/k_{B}}{\eta /\hbar }\times \frac{k_{B}T}{\hbar v/L}%
\times \frac{u_{\mathrm{typ}}}{v}\,,
\end{equation}
where we assumed a typical fluid velocity $u_{\mathrm{typ}}$ and a
characteristic length scale $L$ for the velocity gradients. Thus, the ratio $%
\eta /s$ reveals itself as the key characteristic determining the Reynolds
number, apart from geometrical factors, typical energies and velocities. To
observe turbulence one needs $\mathrm{{Re}>10^{3}\cdots 10^{4}}$ in 3d, and
somewhat higher values in 2d. However, even for lower $\mathrm{{Re}\simeq
10\cdots 10^{2}}$ two dimensional flow in the presence of extended defects
or nano-sized obstacles undergoes complex phase locking phenomena and
chaotic flow~\cite{Saha00}. Applying strong bias fields to graphene, fluid
velocities of the order $u_{\mathrm{typ}}\simeq 0.1 v$ can be achieved~\cite%
{Meric2008}, while still avoiding the onset of non-Ohmic effects. In the
collision-dominated regime of undoped graphene the fluid velocity induced by
a field $E$ scales like $u/v \sim e E\hbar v/(k_BT)^2$. Hence, the Reynolds
number increases as $1/T$ with decreasing $T$, enhancing the tendency
towards turbulence. With flow velocities as estimated above we expect
complex fluid dynamics as in Ref.~\cite{Saha00} already on small length
scales of the order of $L \sim 1\mu\mathrm{m}$. This would constitute a
striking manifestation of the quantum criticality of graphene and could be
relevant for potential nano-electronics applications of this exciting
material.

We thank S. Hartnoll, D. Nelson and S. Sachdev for useful discussions. MM
and JS acknowledge the hospitality of the Aspen Center for Physics. 
The authors were supported by the SNF under grants PA002-113151 and
PP002-118932 (MM), the Ames Laboratory, operated for the US DOE by Iowa
State University under Contract No. DEAC02-07CH11358 (JS), and  DFG grant Fr
2627/1-1 and NSF grant DMR-0757145 (LF).


\begin{thebibliography}{99}
\bibitem{Novoselov04} K. S. Novoselov, A. K. Geim, S. V. Morozov, D. Jiang,
Y. Zhang, S. V. Dubonos, I. V. Grigorieva, and A. A. Firsov, Science \textbf{%
306} 666 (2004).

\bibitem{Novoselov05} K.S. Novoselov, \ A.K. Geim, S.V. Morozov, D. Jiang,
M.I. Katsnelson, I.V. Grigorieva, S.V. Dubonos, A.A. Firsov, Nature \textbf{%
438}, 197 (2005).

\bibitem{Fritz08} L. Fritz, J. Schmalian, M. M\"{u}ller, and S. Sachdev,
Phys. Rev. B \textbf{78}, 085416 (2008).

\bibitem{Sheehy07} D. E. Sheehy and J. Schmalian, Phys. Rev. Lett. \textbf{99%
}, 226803 (2007).

\bibitem{Shuryak03} E. Shuryak, Progress in Particle and Nuclear Physics
\textbf{53}, 273 (2004).

\bibitem{Kovtun03} P. Kovtun, D. T. Son, and A. O. Starinets, Phys. Rev.
Lett. \textbf{94}, 111601 (2005).

\bibitem{counterexample} Y. Kats, P. Petrov, arXiv:0712.0743v3 (2007).
A.~Buchel, R.~C.~Myers, and A.~Sinha, arXiv:0812.2521v1 (2008).

\bibitem{Ioffe60} A.F. Ioffe and A. R. Regel, Prog. Semicond. \textbf{4},
237 (1960).

\bibitem{Mott72} N. Mott, Philosophical Magazine \textbf{26}, 1015-1026
(1972).

\bibitem{ssbook} S.~Sachdev, \emph{Quantum Phase Transitions\/}, Cambridge
University Press, Cambridge (1999).

\bibitem{Herbut08} I. F. Herbut, V. Juri\v{c}i\'{c}, and O. Vafek, Phys.
Rev. Lett. \textbf{100}, 046403 (2008).

\bibitem{Kashuba08} A. B. Kashuba, Phys. Rev. B \textbf{78}, 085415 (2008).

\bibitem{Abrikosov59} A. A. Abrikosov and I. M. Khalatnikov, Rep. Prog.
Phys. \textbf{22}, 329 (1959).

\bibitem{Khalat} I. M. Khalatnikov, An Introduction to the Theory of
Superfluidity, (Benjamin, New York, 1965).

\bibitem{Bruun07} G. M. Bruun and H. Smith, Phys. Rev. A \textbf{75}, 043612
(2007).

\bibitem{Schafer07} T. Sch\"{a}fer, Phys. Rev. A \textbf{76}, 063618 (2007).

\bibitem{Vignale97} G. Vignale, C. A. Ullrich, and S. Conti, Phys. Rev.
Lett. \textbf{79}, 4878 (1997).

\bibitem{Steinberg58} M. S. Steinberg, Phys. Rev. \textbf{109} 1486 (1958).

\bibitem{Weinberg71} S. Weinberg, The Astrophysical Journal \textbf{168},
175 (1971).

\bibitem{Wallace47} P. R. Wallace, Phys. Rev. \textbf{71}, 622 (1947).

\bibitem{Gonzalez99} J. Gonz\'{a}lez, F. Guinea, and M. A. H. Vozmediano,
Nucl. Phys. B \textbf{424}, 595 (1994); Phys. Rev. B \textbf{59}, R2474
(1999).

\bibitem{Ye98} J.~Ye and S.~Sachdev, Phys. Rev. Lett. \textbf{80}, 5409
(1998).

\bibitem{gorbar} E. V. Gorbar, V. P. Gusynin, V. A. Miransky, and I. A.
Shovkovy, Phys. Rev. B \textbf{66}, 045108 (2002).

\bibitem{Son07} D.~T.~Son, Phys. Rev. B \textbf{75}, 235423 (2007).

\bibitem{Damle97} K. Damle and S. Sachdev, Phys. Rev. B \textbf{56}, 8714
(1997)

\bibitem{yaffe00} P.~Arnold, G.~D.~Moore and L.~G.~Yaffe, JHEP \textbf{11},
001 (2000).

\bibitem{ziman} J.~M.~Ziman, \emph{Electrons and Phonons\/}, Oxford
University Press, Oxford (1960), Chapter 7.

\bibitem{Foster2009} M. S. Foster and I. L. Aleiner, Phys. Rev. B \textbf{79}%
, 085415 (2009).

\bibitem{Abanin08} D. A. Abanin and L. S. Levitov, Phys. Rev. B \textbf{78},
035416 (2008).

\bibitem{Fisher88} M.P. A. Fisher and G. Grinstein, Phys. Rev. Lett. \textbf{%
60}, 208 (1988), I. Herbut, Phys. Rev. Lett. \textbf{87}, 137004 (2001).

\bibitem{Herbut06} I. F. Herbut, Phys. Rev. Lett. \textbf{97}, 146401 (2006).

\bibitem{Saha00} A. K. Saha, K. Muralidhar, and G. Biswas, Journal of
Engineering Mechanics-ASCE, \textbf{126}, 523 (2000).

\bibitem{Meric2008} I. Meric, M. Y. Han, A. F. Young, B. Ozyilmaz, P. Kim,
and K. L. Shepard, Nature Nanotechnology \textbf{3}, 654 (2008).
\end{thebibliography}
\end{document}